%% file: NoisestrengthRII.tex
\documentclass[aps,pre,twocolumn,superscriptaddress,showpacs]{revtex4-1}

\usepackage{amssymb}
\usepackage{epsfig}
\usepackage{amsmath}
\usepackage{times}
\usepackage{bm}
\usepackage{nicefrac}
\usepackage{tensor}

\usepackage{epstopdf} 

\usepackage{color}
\definecolor{bluecolor}{rgb}{0,0.,1.}

\definecolor{redcolor}{rgb}{.7,0.,0.}

\begin{document}

\title{Stochastic perturbations in open chaotic systems: random versus noisy maps}

\author{Tam\'{a}s B\'{o}dai}
\affiliation{Klimacampus, Institute of Meteorology, University of Hamburg, Grindelberg 5, Hamburg, D-20144, Germany}
\affiliation{Max Planck Institute for the Physics of Complex
Systems, N\"othnitzer Str. 38, 01187 Dresden, Germany}

\author{Eduardo G. Altmann}
\affiliation{Max Planck Institute for the Physics of Complex
Systems, N\"othnitzer Str. 38, 01187 Dresden, Germany}

\author{Antonio Endler}
\affiliation{Instituto de F\'isica - UFRGS, CP 15051, 91501-970 -
Porto Alegre - RS, Brazil}

\begin{abstract}
We investigate the effects of random perturbations on fully chaotic open systems. Perturbations can be applied to each trajectory independently (white noise) or simultaneously to all trajectories (random map). We compare these two scenarios by generalizing the theory of open chaotic systems and introducing a time-dependent conditionally-map-invariant measure. For the same perturbation strength we show that the escape rate of the random map is always larger than that of the noisy map. In random maps we show that the escape rate $\kappa$ and dimensions $D$ of the relevant fractal sets often depend nonmonotonically on the intensity of the random perturbation. We discuss the accuracy (bias) and precision (variance) of finite-size estimators of $\kappa$ and $D$, and show that the improvement of the precision of the estimations with the number of trajectories $N$ is extremely slow ($\propto 1/\ln N$). We also argue that the finite-size $D$ estimators are typically biased. General theoretical results are combined with {analytical calculations and} numerical simulations in area-preserving baker maps.  
\end{abstract}
\date{\today}

\maketitle

\section{Introduction}

External perturbations affect almost any observation to be made and are usually modeled by simple stochastic processes~\cite{Kapitaniak:1990}. In this paper we are interested in stochastic perturbations in open chaotic maps, i.e., discrete-time systems, which exhibit a transiently chaotic dynamics. Such systems appear in a variety of physical situations (scattering, planetary astronomy, chemical reactions, fluid dynamics, {environmental sciences}, etc.){~\cite{LT:2011}.}

In an {\em ensemble}-based framework, there are two different ways of introducing perturbations, for which the following common terminology applies~\cite{RGO:1990}: 
\begin{itemize}

\item[] {\bf Noisy map:} perturbations are applied {\em independently} to each trajectory.

\item[] {\bf Random map:} the {\em same} perturbation is applied to all trajectories simultaneously.

\end{itemize}

{
Both the noisy- and random-map pictures  appear in numerous {\em physical} systems. In fluid dynamics, molecular diffusion is an example for physical processes that can be modeled by noisy maps, and randomly varying velocity fields affecting fluid advection can be modeled by random maps~\cite{YOC:1991,SO:1993,NOA:1996,JOAY:1997,NT:1998,KTMG:2004}. 
Random map models of fluid dynamics are used whenever a 2-dimensional velocity field shows a non-trivial time-dependence (e.g. when fluid vortices perform complicated movements~\cite{NT:1998}) and have been used to explain experimental observations of fractal spatial patterns of floating particles on the surface of a 3-dimensional fluid~\cite{SO:1993}. Such fractal patterns in random maps have been shown to enhance biological and chemical reactions taking place in fluids, a problem of great interest for the spreading of pollutants in the atmosphere and for the dynamics of plankton in the sea (see Ref.~\cite{KTMG:2004} and references therein). 
In climate and weather models, physical processes on a subgrid level are typically represented by closure relations and parametrizations of relevant diffusion processes, which correspond to the noisy map picture. 
Stochastically parametrized models are a subject of great recent interest because of their potential to improve modeling power and so prediction skills~\cite{PW:2010,CSG:2011}.  
On the other hand, external forcings, e.g. solar irradiation (possibly modulated by major volcano eruptions or anthropogenic CO$_2$ emission), would affect possible weather evolution scenarios the same way, which correspond to the random map picture.~\cite{BT:2012a}. Another situation where the random map picture applies is wave front propagation through randomly structured media, e.g. in underwater acoustics~\cite{Flatte:1979,BW:2011}.} {More generally, noisy maps appear typically in spatially extended systems when {\em microscopic} sources of stochasticity are present, while the random map picture appears when some {\em macroscopic} forcing affects all trajectories simultaneously. From another point of view, noisy maps apply when {\em repeated} experiments with single trajectories are performed, while the random map picture applies when we are interested in the expected outcome of a {\em single} experiment with a fixed realization of the perturbation.  
}

The above distinction can be motivated also from a {\em predictability} point of view, whereby noisy maps describe models with uncertainties, and random maps describe models which are perturbed by an {\em a priori} known process. {Here we are concerned with the predictability of a {\em typical} trajectory, which is arbitrarily chosen from an ensemble, and the measures of predictability will be defined as averages over this ensemble. In the well-studied case of {\em dissipative closed} systems the ensemble {at any time $t$} is taken to be constituted by trajectories which are arbitrarily initialized in the infinitely distant past, $t_0\to-\infty$. In the random map framework this ensemble is referred to as a random or {\em snapshot attractor}, which is a fractal set if the trajectories are chaotic~\cite{RGO:1990}.} A remarkable property of the snapshot attractor is that its geometry and the measure supported by it are changing continuously in time, but its fractal dimension is constant \cite{LY:1988,RGO:1990,JOAY:1997}. {However, e.g. the finite-time average maximal Lyapunov exponent, quantifying the finite-lead-time predictability of the typical trajectory, is time-dependent \cite{BKT:2011}. On the other hand, in the noisy map framework the average point-wise prediction error at some finite lead time cannot be arbitrarily reduced by improving the precision of the initial conditions.} The random map picture may {thus} seem to be in stark contrast to the noisy dynamics {from a predictability point of view}, however, {from a more fundamental point of view} it has been shown that fractal snapshot attractors constitute building blocks of the noisy stationary attractor~\cite{BKT:2011}. 

In autonomous chaotic {\em open} systems the density of trajectories surviving for a long time inside the system, not leaving a window of observation, decays exponentially, and it is distributed according to the so-called {\em conditionally invariant} measure~\cite{PY:1979,DY:2006,LT:2011}. In these systems, predictability concerns {whether we can foretell} the route of escape of a trajectory when there are a number of options for that. A measure of this predictability of the typical, i.e., arbitrarily initialized trajectory in the observational window is the {\em uncertainty exponent}. {The latter is the scaling exponent of the fraction of predictable trajectories with respect to the precision of the initial conditions~\cite{TG:2006}.} {For random maps} the uncertainty exponent is thus trivially related to the fractal dimension of the boundary of basins from which the trajectories escape through different routes. The fractal scaling of the basin boundary is time-dependent at finite, practically accessible scales, and so predictability (or the `rate' at which it can be improved) is also time-dependent. {For noisy maps the basin boundary is space-filling, and so, similarly as in the case of closed systems, predictability cannot be improved at all.}  

In this paper we concentrate on simple (fully hyperbolic) chaotic open systems and simple stochastic perturbation processes, and focus on the effects of the perturbation strength and on the comparison between noisy and random maps. We combine the concepts of conditionally invariant~\cite{LT:2011} and time-dependent but {\em map-invariant}~\cite{Arnold:1998,BKT:2011,ASY:2005} measures, and argue that the trajectories in the random open maps are distributed according to a time-dependent {conditionally-map-invariant measure (to be clarified below)}. Based on this formalism we obtain that, for the same stochastic perturbation process, the escape rate~$\kappa$ of the random map is always smaller than that of the noisy map. We also investigate the dimension~$D$ of the relevant fractal sets of random open maps, and we discuss the accuracy and precision of finite-size estimators of~$\kappa$ and $D$. Under conditions when {\em noise-enhanced trapping} is observed, i.e., when the average life-time of trajectories is `constructively' increased by noise~\cite{franaszek,Reimann:1996,AE:2010}, we find numerical evidence that both $\kappa$ and $D$ of the associated random map also show a nonmonotonic dependence on the perturbation strength.

The paper is organized as follows. In Sec. \ref{sec:cmeasure} the theory of open systems is generalized for random maps, introducing a time-dependent version of the conditionally invariant measure. In Sec. \ref{sec:escrate} we provide general relations for the escape rate in autonomous, random, and noisy maps, followed by the discussion of finite-size estimators. The analogous investigation for the fractal dimension appears in Sec. \ref{sec:dim}. Finally, our main conclusions are summarized in Sec. \ref{sec:conclusions}.

\section{Measures of open maps}\label{sec:cmeasure}

We consider the temporal evolution over $t=0, \ldots, T$ of an ensemble of $n=1, \ldots, N$ trajectories under the action of the map $\vec{x}_{t+1}=f_t(\vec{x}_t)$ in a $d$-dimensional phase space $\vec{x} \in X$, {when on each iteration with respect to $t$ the map $f_t$ is chosen from an ensemble according to some probability distribution~\cite{JOAY:1997}.} We assume that {members of this ensemble are} invertible, i.e., $x_{t}=f_t^{-1}(x_{t+1})$, open, and fully chaotic (to be clarified below). {Equivalently, we can say that the mapping rule $f$} depends on a control parameter $a$, and we consider perturbations around a fixed value $a_*$ as $a=a_t=a_*+\delta \xi_{t}$, where $\delta$ is the strength of the perturbation, and $\xi_t$'s are independent identically distributed random variables (e.g. Gaussian with zero mean and unit variance), which in general vary across different trajectories $n$ (but do not depend on $\vec{x}_n$). Altogether, the dynamics of the ensemble is written as:
\begin{equation}\label{eq:pertdmap}
 \vec{x}_{t+1,n} = f_{t,n}(\vec{x}_{t,n}) \equiv f(\vec{x}_{t,n},\delta\xi_{t,n}).
\end{equation}
According {to} the theory of open maps \cite{LT:2011,Meiss:1997}, for $t \rightarrow \pm \infty$ almost every trajectory leave a finite region of the phase space $\varGamma \subset X$ in which they exhibit some nontrivial dynamics. A central quantity in our analysis will be the probability density function $\rho(\vec{x},t)$ of {\em surviving} trajectories in $\vec{x} \in \varGamma$ up to time $t$, which is obtained by dividing the number of trajectories in an $\varepsilon$-neighborhood of $\vec{x}$ by the total number of surviving trajectories $N(t)$ (in the limit of $\varepsilon \rightarrow 0$ when $N(0) \rightarrow \infty$). While $N(t)/N(0) \rightarrow 0$ for $t \rightarrow \infty$, the normalized density $\rho(\vec{x},t)$ may approach a nontrivial density and be used to define a {\em measure} ($d\mu = \rho dV_X$, where $dV_X$ denotes a phase space volume element in $X$). Next we discuss in detail the properties of this measure in the cases of autonomous-, random-, and noisy maps.

\subsection{Autonomous maps}\label{sec:measureautmaps}

In the unperturbed case, $\delta=0$ in Eq.~(\ref{eq:pertdmap}), the map is autonomous, and the following results are known from transient chaos theory~\cite{LT:2011}: the dynamics is governed by a time-invariant nonattracting chaotic set in $\varGamma$, also called a {\em chaotic saddle}, which is composed of the points that do not leave $\varGamma$ under the action of the mapping (\ref{eq:pertdmap}) in either direction $t \rightarrow \pm \infty$~\cite{noreturn}. For fully chaotic maps this is a zero measure fractal set, {lying at the intersection of its stable and unstable manifolds, which latter sets are composed by points within $\varGamma$ that never leave $\varGamma$ for $t\rightarrow \infty$ and for $t\rightarrow -\infty$, respectively.} The normalized density $\rho(\vec{x},t)$ converges to a well defined stationary density $\rho(\vec{x})$ for $t \rightarrow \infty$. The measure $\mu$ associated with $\rho(\vec{x})$ is said to be {conditionally invariant} (in brief c-measure and c-density, respectively), because for any set $A\subset\varGamma$ it obeys the following relation~\cite{PY:1979,DY:2006}:
\begin{equation}\label{eq.muc}
\mu(A)=\frac{\mu(f^{-1}(A)}{\mu(f^{-1}(\varGamma))},
\end{equation}
where $f^{-1}(\varGamma) \subset \varGamma$ corresponds to the set of points that do not escape $\varGamma$ over one iteration of $f$. Because of a constant rescaling given by the denominator in Eq.~(\ref{eq.muc}), c-measures are {\em not} invariant under the map $f$, i.e., not $f$- or map-invariant~\cite{ASY:2005}. For clarity, we can refer to them as conditionally-map-invariant. C-measures of autonomous maps are time-invariant, however. The c-measure associated with $\rho(\vec{x})$ is a probability measure~\cite{ASY:2005}, indicating the chance of finding a typical trajectory in a particular area of phase space, provided that it has not escaped until time $t$.

\subsection{Random maps}\label{sec:measurerandmaps}

Consider choosing a random sequence of maps $f_t$ by varying the parameter $a=a_t=a_*+\delta \xi_t$: $f_t(\vec{x}_{t}) \equiv f(\vec{x}_{t};\delta\xi_{t})$. At each time $t$ applying the same perturbation $\xi_t$ to all $N$ trajectories, $\xi_{t,n}=\xi_t$ in Eq.~(\ref{eq:pertdmap}), corresponds to the {random map} approach. The sequence of random perturbations can be indexed by the realization $r$ as $\xi_{t,r}$, with which we have different realizations of the sequence of random maps: $f_{t,r}$. For a fixed realization, we can again consider the set of points, initialized at a particular time $t=t_*$, that never escape $\varGamma$ for $t\rightarrow \pm \infty$. This set is called a snapshot saddle~\cite{LT:2011}, whose geometry generally changes with time $t$. Its unstable manifold at time $t$, from which points never escape $\varGamma$ for $t\rightarrow -\infty$, can be seen as the open map counterpart of snapshot attractors of closed random maps~\cite{NOA:1996}. {This way the snapshot attractor is said to be defined in a {\em pullback} sense as the set which is approximated by identically perturbed trajectories initialized in the infinitely distant past, and more recently it has been referred to as a pullback attractor \cite{Arnold:1998}. The constituent} trajectories are distributed according to a time-dependent sample measure~\cite{CSG:2011}, which is a generalization of the Sinai-Ruelle-Bowen (SRB) measure of the autonomous case~\cite{CSG:2011,RGO:1990}. The unstable manifold of the snapshot saddle, too, depends on the entire history of $\xi_{t,r}$ over $t\in(-\infty,t_*]$ -- whereas the snapshot saddle itself depends also on the future $t\in[t_*,\infty)$.

For an ensemble of trajectories initialized at $t=0$, later at some $t \gg 0$ the normalized density of surviving trajectories $\rho(\vec{x},t)$ will be concentrated around the unstable manifold belonging to the time $t$ snapshot saddle, distributed approximately according to a time- and realization-dependent c-density $\hat{\rho}_{t,r}(\vec{x})$. The associated time-dependent generalization of c-measures supported by the unstable manifold obey the following relation:
\begin{equation}\label{eq:cmeasure}
  \hat{\mu}_{t+1,r}(A) = \frac{\hat{\mu}_{t,r}(f^{-1}_{t,r}(A))}{\hat{\mu}_{t,r}(f^{-1}_{t,r}(\varGamma))}.
\end{equation}
We say that the measure $\hat{\mu}_{t,r}$ is conditionally-map-invariant, but it is not time-invariant.

\subsection{Noisy maps}\label{sec:measurenoisymaps}

The physical picture for noisy maps is provided by molecular diffusion, in which case random perturbations act independently on each particle. In terms of the dynamics described by Eq.~(\ref{eq:pertdmap}), this means that $\xi_{t,n}$ and $\xi_{t,n'}$ are independent for any pair of $n\neq n'$. From the point of view of the random maps, the noisy map corresponds to combining the $N$ trajectories of all $R$ realization, with $R,N \rightarrow \infty$. In case of attractors, this corresponds to combining the snapshot attractors to build up the so-called fuzzy attractor~\cite{BKT:2011}, and the natural measure supported by the fuzzy attractor is the average of those supported by the snapshot attractors: $\tilde{\mu}^{att}(A) = \langle \hat{\mu}_{t,r}^{att}(A)\rangle_r=\langle \hat{\mu}_{t,r}^{att}(A)\rangle_t$ \cite{CSG:2011,Arnold:1998}. This naturally extends to the case of open maps, where the c-measure~$\tilde{\mu}$ of any set $A\subset \varGamma$ of the noisy maps is given by:
\begin{equation}\label{eq:avemuc}
  \tilde{\mu}(A) = \langle \hat{\mu}_{t,r}(A)\rangle_r = \langle \hat{\mu}_{t,r}(A) \rangle_t,
\end{equation}
where the last equality is guaranteed by the ergodicity of $\xi$ and shows that $\tilde{\mu}$ is naturally time-invariant. This means that the normalized density of surviving trajectories in the noisy map $\tilde{\rho}(\vec{x},t)$ converges $\tilde{\rho}(\vec{x})$ for $t \rightarrow \infty$, where $\tilde{\rho}=\langle \hat{\rho}_{t,r} \rangle_r = \langle \hat{\rho}_{t,r} \rangle_t$ {is independent of time or realization}. Note that we use the following notation for averaging with respect to, e.g., realizations: $$\langle\bullet\rangle_r=\lim_{R\to\infty}\langle\bullet\rangle_{r=1}^R=\lim_{R\to\infty}\frac{1}{R}\sum_{r=1}^R\bullet$$

A summary of the relevant measures mentioned in this section is given in Tab.~\ref{tab.measures}. In the remainder of this paper we discuss two fundamental quantities of the dynamics, the escape rate~$\kappa$ and the dimensions~$D$ of the relevant fractal sets. We are mainly interested in comparing results observed in random maps to the corresponding noisy maps (for a fixed distribution of $\xi$ and fixed $\delta$), and also compare these two cases to the unperturbed map for increasing values of the perturbation strength $\delta$.    

\input{table-measures.tex}

\section{Escape rate}\label{sec:escrate}

\subsection{General relations}\label{sec:escgeneral}

{In fully chaotic open systems, in which the dynamics is governed by a nonattractctive chaotic set contained by $\varGamma$, the survival probability inside $\varGamma$ for $t \rightarrow \infty$ decays exponentially:}
\begin{equation}\label{eq:survprob}
    P(t) = \lim\limits_{N(0)
\rightarrow \infty} N(t)/N(0) \sim \exp(-\kappa t),
\end{equation}
where $\kappa$ is the escape rate. 
{In the case of Hamiltonian systems deviations from exponential decay appear in the generic case of mixed phase-space systems; see Refs.~\cite{RMG:2010,AE:2010,KKK:2012} for interesting recent investigations on the effects of noise perturbations in this case.}
%
%
In terms of the analysis based on surviving trajectories proposed in Sec.~\ref{sec:cmeasure}, the exponential decay in Eq.~(\ref{eq:survprob}) corresponds to a fixed fraction $\exp({-\kappa})$ of surviving trajectories not escaping after each time step. Considering that the denominator in the right hand side of Eq.~(\ref{eq.muc}) is a normalization factor accounting for the escape of trajectories in one iteration of $f$, one obtains the well-established relation for autonomous maps~\cite{PY:1979}: 
\begin{equation}\label{eq:kappawithmuc}
 \kappa = - \ln \mu(f^{-1}(\varGamma)). 
\end{equation}

In random open maps, the results of Sec.~\ref{sec:cmeasure} show that Eq.~(\ref{eq:kappawithmuc}) can be applied for each realization leading to a time- and realization-dependent single-step escape rate $\hat{\kappa}_{t,r} = -\ln\hat{\mu}_{t,r}(f_{t,r}^{-1}(\varGamma))$. In the spirit of Eq.~(\ref{eq:survprob}), the physically relevant escape rate for a fixed realization $r$ is obtained by aggregating the escapes over time. Therefore, the overall escape rate of the random map~$\hat{\kappa}$ is given by:
\begin{equation}\label{eq:kappamurandommap}
\hat{\kappa}= \langle \hat{\kappa}_{t,r} \rangle_t = \langle \hat{\kappa}_{t,r} \rangle_r = - \langle \ln\hat{\mu}_{t,r}(f_{t,r}^{-1}(\varGamma)) \rangle_r,
\end{equation}
where we used the ergodicity of the random perturbation $\xi$ as in Eq.~(\ref{eq:avemuc}). Ergodicity guarantees that all (typical) realizations of $\xi$ lead to the same escape rate $\hat{\kappa}$ over $t=0,\ldots,T\rightarrow \infty$, and that this value equals the mean obtained over different realizations $r=1,\ldots,R\rightarrow \infty$.

For noisy maps the c-measure is time-invariant and the expression corresponding to Eq.~(\ref{eq:kappawithmuc}) is as follows:
\begin{equation}\label{eq:kappamunoisy}
  \tilde{\kappa} = - \ln\langle\hat{\mu}_{t,r}(f_{t,r}^{-1}(\varGamma))\rangle_r,
\end{equation}
where we used Eq.~(\ref{eq:avemuc}). 

Comparing Eqs. (\ref{eq:kappamurandommap}) and (\ref{eq:kappamunoisy}) we see that the difference stands in the order of taking the average and the logarithm. We now rewrite Eq.~(\ref{eq:kappamurandommap}) as the logarithm of a geometric mean:
$$\hat{\kappa} = \lim_{R\rightarrow\infty} - \ln \left( \prod_{r=1}^R \hat{\mu}_{t,r}(f_{t,r}^{-1}(\varGamma)) \right)^{1/R},$$
which is known to be always smaller {than- or equal to} the arithmetic mean used for $\tilde{\kappa}$ in Eq.~(\ref{eq:kappamunoisy}). With this, we arrive at our first result:
\begin{equation}\label{eq.noiserandom}
\tilde{\kappa} \le \hat{\kappa},
\end{equation}
i.e., the escape rate~$\tilde{\kappa}$ in the noisy-map configuration (perturbation applied independently to each trajectory) is smaller than- or equal to the escape rate~$\hat{\kappa}$ in the random-map configuration (perturbation applied consistently to all trajectories), for the same random process $\xi_t$ and perturbation strength $\delta$. In fact, this inequality is due to the concavity of the {logarithmic function} 
-- the same way as the inequality of the arithmetic and geometric means. Equality is achieved only when $\hat{\mu}_{t,r}(f_{t,r}^{-1}(\varGamma))$ is independent of time. Typically, however, for increasing $\delta$ we expect $\tilde{\kappa}$ to become increasingly smaller than $\hat{\kappa}$.

\subsection{Finite-size estimation}\label{sec:escstimations}

{Important properties of finite-size $S$ estimators $e^{(S)}$ include the bias or {\em accuracy} and the variance or {\em precision}, which are respectively given by the expected value $\langle e^{(S)}_r\rangle_r$ (minus the true value) and variance $var[ e^{(S)}_r]_r$ of a distribution created by an ensemble of realizations of a relevant quantity. Each realization is produced by assigning random values to members of a finite-size set of the relevant quantity (e.g. initial conditions, sequence of perturbations, etc.). If $e^{(S)}_r$ converges to the true value for any $r$ as $S\rightarrow\infty$, then the estimator is said to be {\em consistent}.} If $\langle e^{(S)}_r\rangle_r$ equals the true value for any $S$, then the estimator is said to be {\em unbiased}. When the estimation would involve a finite $R$ number of realizations, e.g. by simply taking the mean over different finite-size estimates, then to work out the improved precision of this estimation we have to consider the combined estimator $e^{(S,R)}\equiv\langle e^{(S)}_r\rangle_{r=1}^R$, and the distribution created by an ensemble of {\em makeups} of $R$ realizations each. With the standard terminology a makeup is then a realization of a group of realizations.

{In this subsection we consider the nontrivial case of the random maps only.} In the previous subsection we saw that for every (typical) realization $r$ the escape rate converges to the same value $\hat{\kappa}=\langle \hat{\kappa}_{t,r}\rangle_t$ in the limit of observation time~$T\rightarrow \infty$. In practice, $T$ is restricted to a maximum value $T_{max}$ due to the finite number of initial conditions $N(0)$, which, according to Eq. (\ref{eq:survprob}), is: 
\begin{equation}\label{eq:Tmax}
  T_{max} \approx \kappa^{-1} \ln N(0).
\end{equation}
Due to the ergodicity of $\xi$, the same $\hat{\kappa}$ is obtained averaging the time and realization-dependent single-step escape rate over different realizations $\hat{\kappa}=\langle \hat{\kappa}_{t,r}\rangle_r$. We thus see that there are two possible strategies to improve the {precision} of estimating $\hat{\kappa}$: (i) increasing $N(0)$ (which, for simplicity we denote hereafter by $N$) or (ii) increasing $R$. In this section we discuss in detail the finite-$T$, -$R$ and -$N$ estimation of $\hat{\kappa}$, as well as the precision of estimation and its scaling with $T$, $R$, and $N$.

\subsubsection{Accuracy of estimation}\label{sec:escestimators}

It is useful to distinguish between two steps in the estimation of $\hat{\kappa}$: the first corresponds to the estimation of $\hat{\kappa}_{t,r}$, the escape rate for a single iterate and for a single realization of the random map, with a finite number of initial conditions $N$, and the second corresponds to the averaging of $\hat{\kappa}_{t,r}$ over a time interval of length $T$ and in turn a number of $R$ different realizations.

The first step applies to the case of autonomous maps as well, and for simplicity we discuss this step in the framework of autonomous maps. The escape rate $\kappa$ is estimated through estimating the measure inside the escape region, according to Eq.~(\ref{eq:kappawithmuc}). After sufficiently long times, the c-measure $\mu(f^{-1}(\varGamma))$ is estimated simply as the fraction of the surviving trajectories with one iteration of the map, i.e., $N(t+1)/N(t)$. With different realizations of the (finite number) of initial conditions we expect $N(t+1)$ to feature a binomial distribution $B(N(t),\mu(f^{-1}(\varGamma)))$, whose mean is $N(t)\mu(f^{-1}(\varGamma))$. This shows that $N(t+1)/N(t)$ is an {unbiased} finite-$N$ estimator of $\mu(f^{-1}(\varGamma))$. However $-\ln N(t+1)/N(t)$ is a {biased} (inaccurate) finite-$N$ estimator of $\kappa$ because the average is performed after taking the logarithm -- applying once more the same reasoning leading to Eq.~(\ref{eq.noiserandom}). In practice, it is important to guarantee that $N(t+1)$ is sufficiently large so that this bias is sufficiently small. Numerically this is not always easy because $N\propto \exp({-\kappa t})$ and time $t$ must also be sufficiently large in order for $N(t+1)/N(t)$ to be a good estimate of $\mu(f^{-1}(\varGamma))$ (convergence of the initial density to the c-measure $\mu$).

Assuming we have accurate estimates of $\hat{\kappa}_{t,r}$, we proceed to the second step and consider the effect of averaging over time $\langle \hat{\kappa}_{t,r} \rangle_t$ or realizations  $\langle \hat{\kappa}_{t,r} \rangle_r$. As argued by Eq.~(\ref{eq:kappamurandommap}), the ergodicity of $\xi$ guarantees that both averages converge to the same value~$\hat{\kappa}$. In practice it is also interesting to consider finite averages performed {\em simultaneously} over $T$ time steps and $R$ realizations, {resulting in the combined estimator}:
\begin{equation}\label{eq:kappaTR}
  \hat{\kappa}^{(T,R)}\equiv \langle\langle \hat{\kappa}_{t,r}\rangle_{t=t'+1}^{t'+T}\rangle_{r=1}^{R}.
\end{equation}
This also converges to $\hat{\kappa}$ for either $T\rightarrow\infty$ or $R\rightarrow\infty$, which makes it a {consistent} estimator. Moreover, the finite-$T$ and finite-$R$ estimations are also unbiased. For this we consider the values of $\hat{\kappa}^{(T,R)}$ obtained for a set of $m=1, \ldots, M\rightarrow\infty$ makeups, each one with (fixed) $R$ realizations and $T$ time steps for the estimation. According to the central limit theorem, the distribution of the estimates with respect to the makeups $m$ will be approximately normal with average equal to $\hat{\kappa}$.

\subsubsection{Precision of estimation and its scaling with $T,R,$ and $N$}\label{sec:escconv}

The variance of the same distribution scales with the number of terms being averaged as follows:
\begin{equation}\label{eq:ftescscaling}
    \sigma_{\langle\hat{\kappa}\rangle}^2 \equiv \langle (\hat{\kappa}^{(T,R)})^2 \rangle_m - \langle \hat{\kappa}^{(T,R)} \rangle_m^2 \sim \frac{\sigma_{\hat{\kappa}}^2}{TR}.
\end{equation}
where $\sigma_{\hat{\kappa}}^2=\sigma_{\langle\hat{\kappa}\rangle}^2(T=1,R=1)=var[\hat{\kappa}_{t,r}]_t=var[\hat{\kappa}_{t,r}]_r$.

We now compare the the two different strategies of improving the precision of estimating $\hat{\kappa}$: (i) $R$ fixed and $T\rightarrow \infty$; and (ii) $T$ fixed and $R\rightarrow \infty$. For a fixed $N$ number of trajectories all trajectories escape in some finite time $T_{max}$ {given by Eq. (\ref{eq:Tmax}). Substituting the latter into Eq. (\ref{eq:ftescscaling})} we arrive at the scaling law:  
\begin{equation}\label{eq:ftescscalingN0}
    \sigma_{\langle\hat{\kappa}\rangle}^2 \propto 1/\ln N.
\end{equation}
The latter indicates that with strategy (i) a steady improvement of the precision of estimates can be achieved by increasing the number of trajectories $N$ with an exponential rate. Thus, the precision can be improved much more effectively using strategy (ii), i.e., by increasing the number of realizations $R$, so that the improvement, according to Eq. (\ref{eq:ftescscaling}), is (inversely) proportional (as opposed to a logarithmic relation) to $R$.

\subsection{Examples}\label{sec:escexamples}

To illustrate aspects of transient chaos in fully chaotic open systems an area-preserving baker map will be analyzed. A general property of the baker map, $(x_{t+1},y_{t+1})=B(x_t,y_t)$, is that the mapping rule $B$ is defined in a piece-wise manner, such as:
\begin{subequations}\label{eq:bakergeneral}
    \begin{eqnarray}
    B_-&=&(x_t/a,ay_t),\ y_t<1/2,\\
    B_+&=&(1+(x_t-1)/a,1+a(y_t-1)),\ y_t>1/2,
    \end{eqnarray}
\end{subequations}
where $a$ is the only free parameter. Trajectories mapped outside the unit square $(x,y)\in [0,1] \times [0,1]$ are considered to have escaped (open boundaries).

\subsubsection*{Example 1: Area-preserving naturally open baker
map}

In this example we obtain an analytic expressions for the escape rate that illustrates inequality~(\ref{eq.noiserandom}). For $a>2$ the map is said to be naturally {open}. The stochastic perturbation is added in $a$ around a fixed value $a_*$ as  $a_t = a_* + \delta \xi_t$, where $a_t\ge2$, and $\xi_t$ is an independent identically distributed (iid) random variable with zero mean and finite variance $\sigma_\xi$. 

First, consider the random map. Starting with the unit square, after $T$ iterations a number of $2^T$ strips of equal width $\Pi_{t=1}^T1/a_t$ remain. The escape rate can be thus obtained
as~\cite{NT:1998,LT:2011}: 
\begin{equation}\label{eq:bakerescrand}
    \begin{split}
    \hat{\kappa} &=
    -\left\langle\ln\frac{2}{a_t}\right\rangle_{t} = \left\langle \ln \frac{a_r}{2} \right\rangle_r \approx \kappa_* -\dfrac{1}{2}\left(\dfrac{\delta \sigma_\xi}{a_*}\right)^2,
    \end{split}
\end{equation}
where  $\kappa_*= \ln (2/a_*)$ is the unperturbed escape rate, ergodicity~(\ref{eq:avemuc}) has been used, and the approximation is obtained as a second order Taylor expansion in $\delta$. It is worth noting that the escape rate decreases with the perturbation intensity, i.e., the trapping is enhanced by the perturbation. A comparison of Eq. (\ref{eq:bakerescrand}) with Eq. (\ref{eq:kappamurandommap}) reveals that the {fraction of surviving trajectories in terms of the c-measure} is $\hat{\mu}_t(f_t^{-1}(\varGamma)) = 2/a_t$. Notice that in this simple example the c-measure depends only on the current value of the perturbation, but not on the complete history. It is a consequence of the fact that at each time step a fraction $1-2/a_t$ of the surviving trajectories escape. The logarithm of the survival probability $\ln P(t)$ can be thought of as a simple random walk (with a drift $\hat{\kappa}$) and therefore the scaling laws discussed in Sec.~\ref{sec:escconv} can be obtained explicitly. The reduction of $\hat{\kappa}$ with $\delta$ (noise-enhanced trapping) can be understood in this case simply as a consequence of the concavity of the logarithmic function $\kappa=\ln(a/2)$.

Next we consider the noisy map, apply Eq.~(\ref{eq:kappamunoisy}), and take the following approximations:
\begin{equation}\label{eq:bakeescnoisy}
    \begin{split}
    \tilde{\kappa} &= -\ln\left\langle\frac{2}{a_r}\right\rangle_r 
\approx -\ln \left(\frac{2}{a_*}\left[1+\left(\frac{\delta \sigma_\xi}{a_*}\right)^2\right]\right) \\ 
&\approx \kappa_* - \left(\frac{\delta \sigma_\xi}{a_*}\right)^2 = \hat{\kappa} - \frac{1}{2} \left(\frac{\delta \sigma_\xi}{a_*}\right)^2 \le \hat{\kappa}.
    \end{split}
\end{equation}
Interestingly, for small perturbations the noise increases the trapping by reducing $\kappa_*$ by twice the amount as in the case of random
maps. {The authors of Ref. \cite{KTMG:2004} find the same quadratic deviation of the mean of the logarithm and the logarithm of the mean of a random variable, for small `strengths of its randomness' $\delta \sigma_\xi/a_*$, corresponding with the second order approximations in our Eqs. (\ref{eq:bakerescrand}) and (\ref{eq:bakeescnoisy}).} Choosing $\xi$ to be uniformly distributed in $[-1,1]$, we can also compute $\hat{\kappa}$ and $\tilde{\kappa}$ exactly using $\langle \ldots \rangle_t=\langle \ldots \rangle_r = \frac{1}{2} \int_{-1}^{1} \ldots d \xi$. For $a_*=2.5$ and $\delta=0.5$ we obtain $\hat{\kappa}=0.2164...$ and $\tilde{\kappa}=0.2096...$.

\subsubsection*{Example 2: Area-preserving closed baker map with a leak}\label{sec:trapping}

In our second example we explore a case in which no simple analytic expressions for $\hat{\mu}_{t,r}$ and $\hat{\kappa}$ exist, but in which numerical results confirm the validity of the scaling $\sigma^2_{\langle \hat{\kappa} \rangle} \propto 1/ T$ (\ref{eq:ftescscaling}), the inequality $\tilde{\kappa} < \hat{\kappa}$ (\ref{eq.noiserandom}), and noise-enhanced trapping for small $\delta$. We start from the area-preserving {closed} baker map, obtained by fixing~$a=2$ in Eq.~(\ref{eq:bakergeneral}), into which we introduce a leak of width $\Delta_x=0.1$ vertically in the middle of the map's phase space. Any trajectory that is mapped into the leak region is considered to have escaped. Differently from the previous example, here the stochastic perturbation is defined to act on the coordinates independently, such that: $x\rightarrow x+\delta\xi_x$ and $y\rightarrow y+\delta\xi_y$, where $\xi_x,\xi_y\in[-1,1]$ are uniformly distributed independent random variables. This way escape out of the unit square is also possible.  

\begin{figure*}[th!]
\centerline{\psfig{figure=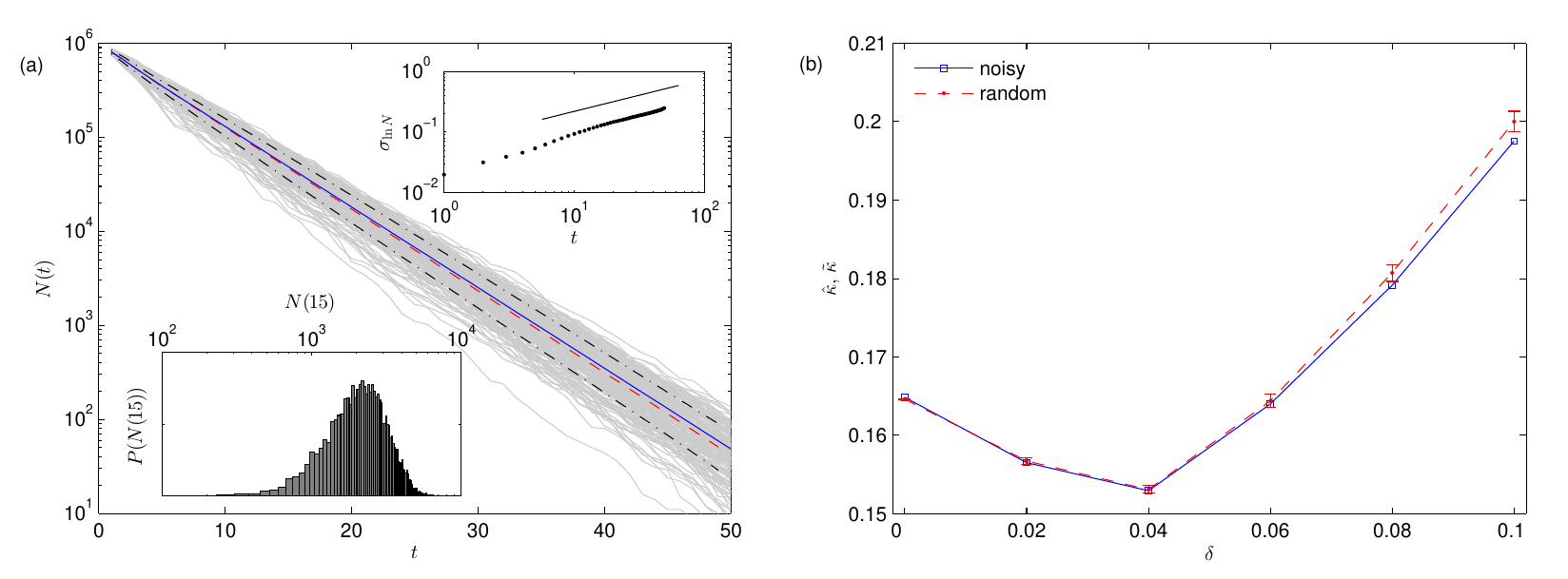,width=\linewidth}}
\caption{(Color online) Difference between the escape rate in the noisy ($\tilde{\kappa}$) and random maps ($\hat{\kappa}$). (a) Number $N(t)$ of surviving trajectories out of $N(0)=10^6$ over time for perturbation strength $\delta=0.1$ in the area-preserving closed {random} baker map with a central vertical leak of width $\Delta_x=0.1$. A number of $R=10^4$ experiments (realizations of $\xi$) were carried out and (200 of them) are shown as gray lines in the backdrop. The lines in the front show: $\langle N \rangle_{r=1}^R$ (blue solid line), corresponding to the noisy map as in Eq. (\ref{eq:kappamunoisy});  $\exp{\langle \ln N \rangle_{r=1}^{R}}$ (red dashed line), corresponding to the random map as in Eq. (\ref{eq:kappamurandommap}); and $\exp(\langle \ln N \rangle_{r=1}^{R} \pm \sigma_{\ln N}/2)$ (dash-dot lines), corresponding to the standard deviation around the expected value in the random map case. Lower inset: histogram of $N(t=15)$; upper inset: scaling of the dispersion $\sigma_{\ln N}$ (half the vertical space between the dash-dot lines) as $t^{1/2}$ (the solid straight line indicates a slope of 1/2). (b) Dependence of $\tilde{\kappa}$ and $\hat{\kappa}$ on the perturbation strength $\delta$. A number of $R=10^4\times\delta$ experiments are done for each $\delta>0$ (and this time for the noisy map all $N(0)=10^6$ trajectories are perturbed independently, in all the $R$ experiments). The escape rates $\tilde{\kappa}$ and $\hat{\kappa}$ were estimated from data from time $t=6$ up to $20$ (giving  $T=14$). In the case of the random map errorbars indicate the standard deviation $\sigma_{\langle\hat{\kappa}\rangle}$. In the case of the noisy map the errorbars $\sigma_{\langle \tilde{\kappa}\rangle}$ are smaller than the square marker. In (a) and (b) data points  are connected  by lines to guide the eye.
} \label{fig:kappavsigvert}
\end{figure*}

In Fig.~\ref{fig:kappavsigvert}~(a) we show the temporal decay of the number of surviving trajectories $N(t)$ for $R=10^4$ different realizations of $\xi_t$ (all $N(0)=10^6$ trajectories are exposed to the {\em same} sequence $\xi_t$) \cite{nohat}. In agreement with Eq.~(\ref{eq:kappamurandommap}), the escape rate of the random map $\hat{\kappa}$ can be obtained by averaging the logarithm of the number of surviving trajectories over different realizations, i.e., $\langle \ln N(t) \rangle_r \sim \ln N(0) - \hat{\kappa}t$. In Fig.~\ref{fig:kappavsigvert}~(a) this slope is shown to be {somewhat} larger than the value $\tilde{\kappa}$ which in principle (for $N(0)\rightarrow\infty$) corresponds to independent perturbations applied to each initial condition, so that $\ln \langle N(t) \rangle_r \sim \ln N(0) -\tilde{\kappa}t$. It is expected that $\tilde{\kappa}$ is obtained here rather inaccurately, because our procedure acts as if $\tilde{\kappa}$ was determined from a small $R=10^4$ number of trajectories, and the value was redundantly replicated-and-averaged $N(0)=10^6$ times. Nevertheless, the results are suitable to indicate the general inequality (\ref{eq.noiserandom}). In the insets of Fig. \ref{fig:kappavsigvert}~(a) we explore the dispersion of $\ln N(t)$, which is shown to follow a $t^{1/2}$ scaling, according to the general rule (\ref{eq:ftescscaling}) and observing Eq. (\ref{eq:survprob}). Notice that proper scaling is attained after a relatively short transient time ($t'$ as in Eq. (\ref{eq:kappaTR})), once the initially uniform densities closely approximate the c-density. The dependence of $\tilde{\kappa}$ ($\square$) and $\hat{\kappa}$ ($\bullet$) on the noise strength $\delta$ is shown in Fig. \ref{fig:kappavsigvert}~(b). The difference between $\tilde{\kappa}$ and $\hat{\kappa}$ steadily increases with the perturbation strength and both curves show a nonmonotonic dependence with a local minimum. This effect has been explained in~\cite{AE:2010} as follows: for small $\delta$  the probability of escape through the leak is reduced $\tilde{\kappa} > \kappa_{\delta=0}$ because the invariant density over the leak is smoothed in comparison with the unperturbed case which has a fractal support; for large noise more and more trajectories escape through the open boundaries, leading to an increase in $\tilde{\kappa}$. {From a different perspective, Ref.~\cite{Klages:2002} develops a framework for understanding a similar nonmonotonic behavior in the diffusion coefficient of a chain of chaotic maps under small noise perturbations. The idea is to consider the dependence of the diffusion coefficient~$\mathcal{D}$ on the perturbed parameter~$a$, and then compute the perturbed diffusion coefficient for small perturbations as an integral of $\mathcal{D}(a)$ over the range of perturbations in $a$ (observing the probability density of the perturbation process). We see that applying this approach to the case of the escape rate~\cite{Reimann:1996}, we obtain the random map escape rate~$\hat{\kappa}$ for {\em Example 1} as given by Eq.~(\ref{eq:bakerescrand}). This approximation is exact only in special cases (like in case of {\em Example 1}) when the escape at time $t$ depends only on the value of~$a$ at time $t$ and it is independent of the values at any time $t'<t$. 
It remains to be shown to what extent this approximation explains non-monotonic dependence of $\hat{\kappa}$ and $\tilde{\kappa}$ on $\delta$ for {\em Example 2} and for chaotic systems more generally.
Overall, here we have shown that $\hat{\kappa}(\delta)$ follows the same general trends as $\tilde{\kappa}(\delta)$, but the noise-enhanced trapping is less} effective due to the inequality (\ref{eq.noiserandom}). In the next section we explore the effect of the random perturbation on the associated fractal dimension.   

\section{Dimensions}\label{sec:dim}

\subsection{General relations}

An important difference between the dimension and escape rate is that the noisy perturbations wash out the fine details of the invariant sets and the (asymptotic) dimensions are not fractional $D=D_{\text{phase space}}$ \cite{BMPG:1985}. Speaking about fractality is thus meaningful only in terms of the random map, and a nontrivial inequality like (\ref{eq.noiserandom}) cannot be established for the dimensions. Here we discuss three different relationships between the dimension and other quantities. {In the next section, subsequently}, estimations of the dimensions will be based on these relationships: {(I) based on box-counting (BC), (II) the Kantz-Grassberger (KG) relationship, and (III) based on the uncertainty exponent ($\alpha$).}

(I:BC) The fractal dimension is defined directly by the relevant {measures} discussed in Sec.~\ref{sec:cmeasure}. The full fractal dimension spectrum $D_q$ of a measure can be computed by applying e.g. a box-counting algorithm. The relevant measures in case of random open maps are the sample measure supported by the random saddle and the c-measure supported by the unstable manifold of the random saddle. The dimension spectrum of the sample measure in the stable direction is identical with that of the c-measure, denoted as $D_q^{(2)}$. Although large scale features of $\hat{\mu}_t$ in phase space, in association with a limited memory of the past, depend on time, its asymptotically fine details are determined by the complete history of the system's evolution. The fractal dimension quantifying the asymptotic scaling of the measure is thus 
constant~\cite{LY:1988}. This too makes the dimension a qualitatively different characteristic number from that of the escape rate $\hat{\kappa}_{t}$, which changes in time together with $\hat{\mu}_t$ [Eq.~(\ref{eq:kappamurandommap})].

(II:KG) For autonomous fully chaotic {2D} open maps the connection between the geometry and the dynamics on the chaotic saddle is established by the pair of the Kantz--Grassberger relations, which reads as follows \cite{KG:1985}:
\begin{subequations}\label{eq:KG}
    \begin{eqnarray}
    D_1^{(1)} &=& 1-\frac{\kappa}{\lambda},\\
    D_1^{(2)} &=& \frac{\lambda-\kappa}{|\lambda'|},
    \end{eqnarray}
\end{subequations}
where $D_1^{(1)}$ and $D_1^{(2)}$ are the partial information dimensions across (along) the stable (unstable) and unstable (stable) manifolds, respectively, and $\lambda$ and $\lambda'$ are the corresponding average positive and negative Lyapunov exponents {\em on} the chaotic saddle, respectively \cite{musamplevsc}. In conservative area-preserving systems $\lambda=-\lambda'$, as a result of which: $D_1^{(1)}=D_1^{(2)}$, {and e.g. $D_q^{(1)}=D_0^{(1)}$ for any $q$}. In case of random maps, similar relations to (\ref{eq:KG}) hold \cite{JOAY:1997}:
\begin{subequations}\label{eq:KGrand}
    \begin{eqnarray}
    \hat{D}_1^{(1)} &=& 1-\frac{\hat{\kappa}}{\hat{\lambda}},\\
    \hat{D}_1^{(2)} &=& \frac{\hat{\lambda}-\hat{\kappa}}{|\hat{\lambda}'|}.
    \end{eqnarray}
\end{subequations}
Analogous to $\hat{\kappa}_{t,r}$ (except for \cite{musamplevsc}), we can define the time- and realization-dependent one-step positive and negative average Lyapunov exponents as ensemble averages, e.g. $\hat{\lambda}_{t,r} = \langle \hat{\lambda}_{t,r,n} \rangle_n$ ($n$ as in Eq. (\ref{eq:pertdmap})), and with this $\hat{\lambda} = \langle \hat{\lambda}_{t,r} \rangle_t = \langle \hat{\lambda}_{t,r} \rangle_r$, where the latter equality is due to the ergodicity of noise. The same applies to $\hat{\lambda'}$. For each trajectory $\hat{\lambda}_{t,r,n} = \ln\hat{y}_{t,r,n}$ and $\hat{\lambda}'_{t,r,n} = \ln\hat{y}'_{t,r,n}$ are defined by $\hat{y}_{t,r,n}$ and $\hat{y}'_{t,r,n}$, which are respectively the stepwise stretching and shrinking rates along the corresponding covariant Lyapunov vectors, i.e., the corresponding manifolds.

(III:$\alpha$) Fractality can be related with the concept of {\em uncertainty}. The latter is measured by the uncertainty exponent $\alpha$, which specifies the scaling of the ratio of the number of uncertain boxes to that of all the boxes with the resolution or box size, such as: $\mathcal{N}_b(\varepsilon)/\mathcal{N}_0(\varepsilon)\propto\varepsilon^{\alpha}$. The certainty of a box is defined so that {\em any} trajectory from it takes the same route of escape. For example in case of the unperturbed baker map, Example 1 of Sec. \ref{sec:trapping}, trajectories can escape from either the left or the right side of the leak. The uncertain boxes for $\varepsilon\rightarrow 0$ {shrink onto} the stable manifold of the chaotic saddle, and thus the uncertainty exponent is related to the Hausdorff dimension as~\cite{TG:2006}:
\begin{equation}\label{eq:ucexpanddim}
D_0^{(1)}=1-\alpha,
\end{equation}
because the number of all boxes in a plane scales as: $\mathcal{N}_0(\varepsilon)\propto\varepsilon^{-2}$. The uncertainty grows with decreasing $\alpha$, which means that it gets more difficult to improve the predictability of the outcome for the typical (randomly chosen) trajectory by increasing the precision of the initial condition. Equation (\ref{eq:ucexpanddim}) indicates that the stronger the fractality the greater the uncertainty, and the same relation holds for noisy as well as random maps. For noisy maps predictability cannot be arbitrarily improved by improving the precision in the initial conditions, consistent with $\tilde{\alpha}=0$ and $\tilde{D}_0^{(1)}=1$. If the perturbation history is known, then the random map framework is relevant (in which case the term `random' is rather misleading), fractality is resolved ($\hat{D}_0^{(1)}<1$), and predictability can be improved ($\hat{\alpha}>0$), similarly as in the unperturbed case.

\subsection{Finite-size estimations}

Different algorithms for the computation or estimation of the fractal dimension based on the relations of the previous section (I,II,III) are commonly used. Here we discuss the accuracy (bias and consistency) and precision (variance or spread) of the estimations based on these algorithms when applied to random maps. As discussed before [see Eq. (\ref{eq:Tmax})], with finite $N$ number of trajectories numerical experiments are limited to a maximum time $T_{max}$~\cite{morefundamental}, and  estimations are based preferably on averaging over different experiments, that is, realizations~$R$ of the perturbations, labeled as strategy (ii) in Sec.~\ref{sec:escstimations}.

\subsubsection{List of estimators}\label{sec:averaging}

(I:BC) Most naturally the fractal dimension is calculated by a direct estimation of the scaling of $\ln \mathcal{N}_b(\varepsilon)$ with $-\ln \varepsilon$ (or that of the information for the information dimension). $\mathcal{N}_b$ is approximated by the count of boxes in a regular rectangular grid that contain at least one point out of finite $N(T<T_{max})$ points that represent e.g. the unstable manifold. The scaling line is fitted by a straight line over a finite $\varepsilon$-range, up to a minimal box size $\varepsilon_*$, whose slope estimates the dimension $\hat{D}_{BC,r}^{(2,\varepsilon_*)}$. The actual estimator is then defined as the average over $R$ realizations:
\begin{equation}\label{eq.DBC}
\hat{D}_{BC} \equiv \hat{D}_{BC}^{(2,\varepsilon_*,R)} = \langle \hat{D}_{BC,r}^{(2,\varepsilon_*)} \rangle_{r=1}^R
\end{equation}

(II:KG.a) Another type of estimator is based on the KG relations~(\ref{eq:KGrand}). For simplicity, we focus only on the second of these relations (the same conclusions apply to the other option). Similarly as for $\hat{D}_{BC}$, we define the estimator as an average of the single-realization KG estimator over $R$ realizations:
\begin{equation}\label{eq:DKGa}
    \hat{D}_{KGa} \equiv \hat{D}_{KGa}^{(2,T,R)} =
\langle \hat{D}_{KGa,r}^{(2,T)} \rangle_{r=1}^{R} = \left\langle\frac{\hat{\lambda}^{(T)}_r - \hat{\kappa}^{(T)}_r}{|\hat{\lambda}'^{(T)}_r|}\right\rangle_{r=1}^{R},
\end{equation}
where $\hat{\kappa}^{(T)}_r=\langle\hat{\kappa}_{t,r}\rangle_{t=t'}^{t'+T}$ ($t'>1$ for well-approximating the measure) is a version of $\hat{\kappa}^{(T,R)}$ of Eq. (\ref{eq:kappaTR}), without averaging with respect to the realization, and similarly e.g. $\hat{\lambda}^{(T)}_r=\langle\hat{\lambda}_{t,r}\rangle_{t=t'}^{t'+T}$~\cite{twoexperiments}.

(II:KG.b) An alternative KG estimator could be defined by changing the order of division and averaging:
\begin{equation}\label{eq:DKGb}
    \hat{D}_{KGb} \equiv \hat{D}_{KGb}^{(2,T,R)}  = \frac{\hat{\lambda}^{(T,R)} - \hat{\kappa}^{(T,R)}}{|\hat{\lambda}'^{(T,R)}|} = \frac{\langle
      \hat{\lambda}^{(T)}_r - \hat{\kappa}^{(T)}_r\rangle_{r=1}^{R}}{\langle |\hat{\lambda}'^{(T)}_r|\rangle_{r=1}^{R}},
\end{equation}
where the last equality follows from the convention (\ref{eq:kappaTR}). Note that here we do not have a single-realization estimator for the dimension, but only for the escape rate and the Lyapunov exponents.

(III:$\alpha$) The stable manifold is contained by the basin boundary, which separates regions of initial conditions from which trajectories take different routes of escape. Thus, the number of uncertain boxes can be approximated also by an approximate finite-$\varepsilon_{min}$ resolution survey of the basin boundary. Trajectories are initialized on a regular rectangular array of points, and all of them will escape one way or another in a finite time. The route of escape will assign logical values to the grid points. The scale is varied by a rectangular grouping of $k^2$ grid points, so that $\varepsilon = k\varepsilon_{min}$. A box of size $\varepsilon$ is certain if all the $k^2$ grid points are assigned the same logical value. This way we have another means of box-counting estimation of either the uncertainty exponent, or, according to relation (\ref{eq:ucexpanddim}), the Hausdorff dimension of the stable manifold: 
\begin{equation}\label{eq:Dalpha}
    \hat{D}_{\alpha} \equiv \hat{D}_{\alpha}^{(1,\varepsilon_{min},R)} = \langle \hat{D}_{\alpha,r}^{(1,\varepsilon_{min})} \rangle_{r=1}^R = 1-\langle \alpha^{\varepsilon_{min}}_r \rangle_{r=1}^R.
\end{equation}

\subsubsection{Accuracy of estimation}

Similarly to the discussion of the escape rate, we first discuss the accuracy of the finite-time estimation, and then the accuracy in relation with averaging over realizations. A single-step and single-realization measure of the dimension, akin to $\hat{\kappa}_{t,r}$, has not been defined, because the true dimension is constant. However, the single-realization box-counting estimate $\hat{D}_{BC,r}^{(2,\varepsilon_*)}$ is found to be time- and realization-dependent. This can be explained with its relationship to the finite-time single-realization estimator $\hat{D}_{KGa,r}^{(2,T)}$ as follows. First, in terms of the baker map, which is a paradigmatic example for e.g. the fractal structure of chaotic attractors or saddles, a number of $T$ iterations determine the scaling of geometry down to a size $\varepsilon \propto |\lambda'|^T$ (see Example 2 of Sec. \ref{sec:dimexamples}). Second, with a finite $N$ number of points to evaluate $\hat{D}_{BC,r}^{(2,\varepsilon_*)}$, fractal scaling holds down to a scale $\varepsilon_*$ related to $N$ as: $\ln1/\varepsilon_*\propto\ln N$. These two points suggest that with finite $N$, the system has a memory of 
\begin{equation}\label{eq:memory}
 T \propto \ln1/\varepsilon_* \propto \ln N
\end{equation}
steps back into the past. Indeed, we have found numerically $\hat{D}_{BC,r}^{(2,\varepsilon_*)}$ to correlate with $\hat{D}_{KGa,r}^{(2,T)}$ for an appropriate $T$. Thus, the time-dependence of the latter can be related to the time-dependent finite-memory/scale properties of the measure. In the limit $N\rightarrow\infty$, the entire past determines the dimension, and both $\hat{D}_{BC,r}^{(2,\varepsilon_*)}$ and $\hat{D}_{KGa,r}^{(2,T)}$ converge to the true value $D^{(2)}_1$ as $\ln1/\varepsilon_*$ and $T\rightarrow\infty$, for any realization. For finite $T$ and $N$, the accuracy of $\hat{D}_{KGa,r}^{(2,T)}$ depends on the accuracy of $\hat{\kappa}^{(T)}_r$ and $\hat{\lambda}^{(T)}_r$. It has been argued in Sec. \ref{sec:escestimators} that $\hat{\kappa}^{(T)}_r$ is biased, but it can be effectively reduced by choosing $N$ large enough. The situation with $\hat{\lambda}^{(T)}_r$ is similar, and therefore, so it is with $\hat{D}_{KGa,r}^{(2,T)}$. As for $\hat{D}_{BC,r}^{(2,\varepsilon_*)}$, some box-counting algorithms has been reported in~\cite{HW:1993} to be negatively biased, which agrees with our numerical experiences. Unlike the bias of $\hat{D}_{KGa,r}^{(2,T)}$, the bias of $\hat{D}_{BC,r}^{(2,\varepsilon_*)}$ may not be insignificant relative to other components of the bias of the total estimator $\hat{D}_{BC}^{(2,\varepsilon_*,R)}$.

Fitting the approximate scaling line of $\mathcal{N}_b(\varepsilon)/\mathcal{N}_0(\varepsilon)$ to calculate the dimension through the uncertainty exponent, according to Eq. (\ref{eq:Dalpha}), leads to a different bias than what we have if directly the scaling line of $\mathcal{N}_b(\varepsilon)$ is fitted.

Next, we focus on the second step, i.e., the averaging over different realizations. We start with comparing $D_{KGa}$ and $D_{KGb}$. First we recall our results from Sec.~\ref{sec:escestimators} that $\hat{\kappa}^{(T,R)}$ is an unbiased estimator of $\hat{\kappa}$ (assuming that $N\rightarrow\infty$). We can argue that the same applies to $\hat{\lambda}^{(T,R)}$. Therefore, $\hat{D}_{KGb}\rightarrow \hat{D}$ as $R \rightarrow \infty$, and  we conclude that $\hat{D}_{KGb}$ is a consistent estimator. Now we notice that $D_{KGb}$ and $D_{KGa}$ will typically lead to different values because the ratio of the averages is different from the average of the ratios. For the case of iid random variables we have: $\langle x/y \rangle = \langle x \rangle \langle 1/y \rangle > \langle x \rangle / \langle y \rangle$. (The inequality holds also for dependent $x$ and $y$.) {Therefore we conclude} that $D_{KGa}$ is not a consistent estimator, and it always {\em overestimates} the true value. The bias of these estimators at finite $R$ can be determined by considering the average value of $D_{KGa}$ and $D_{KGb}$ over different makeups (as in Sec.~\ref{sec:escestimators}). In this case, both estimators would correspond to the average of ratios instead of the ratio of averages. Therefore, we conclude that both $D_{KGa}$ and $D_{KGb}$ are biased estimators. 

As for $\hat{D}_{BC}$, even if we do not have an analytical model of it, the correlation between $\hat{D}_{BC,r}^{(2,\varepsilon_*)}$ and $\hat{D}_{KGa,r}^{(2,T)}$ suggests that $\hat{D}_{BC}$ should also be expected to be biased (beside the effect reported in~\cite{HW:1993}). Because of its similar nature, the same can be said about $\hat{D}_{\alpha}$. 

In the example above the source of the bias was taking the arithmetic mean $\langle \ldots \rangle_{r=1}^{R}$, and it is natural to ask ourselves whether other approaches could lead to a better estimate. {In this regard e.g. the harmonic mean, $(\langle \hat{D}_r^{-1} \rangle_{r=1}^{R})^{-1}$, is not expected to perform better when the numerator is not constant.} In case of iid random variables $x$ and $y$, again with a reference to the numerator and denominator of $\hat{D}_{KGa,r}^{(2,T)}$, the {\em harmonic mean} {\em underestimates} the true value: $\langle y/x\rangle^{-1} = \langle y\rangle^{-1}\langle x^{-1}\rangle^{-1} < \langle x\rangle/\langle y\rangle$. When $\sigma_x^2/\langle x \rangle^2$ is much greater (smaller) than $\sigma_y^2/\langle y \rangle^2$, the arithmetic (harmonic) mean gives the better approximation, since we find that e.g. $\langle y \rangle/\langle y^{-1} \rangle^{-1}-1 \propto \sigma_y^2/\langle y \rangle^2$. It is easy to verify that for equal values of these, the harmonic mean yields smaller bias. 

To the end of creating an unbiased $\hat{D}_{KGa}^{(2,T,R)}$-type estimator, an appropriate generalized $f$-mean 
can be found only when the numerator and denominator are interrelated in a special way. As an example, when the Jacobian $|\hat{J}_t|=\hat{y}_t\hat{y}'_t$ of a closed dissipative system is constant, as in case of the fluid flow in \cite{NOA:1996}, the appropriate generalized $f$-mean can be characterized by the form: $f(\hat{D}_{L,r}^{(2,T)}) = \hat{D}_{L,r}^{(2,T)}/(\hat{D}_{L,r}^{(2,T)}-1)$. In the latter $\hat{D}_{L,r}^{(2,T)}$ is the Lyapunov dimension that can be obtained from $\hat{D}_{KGa,r}^{(2,T)}$ when $\kappa\rightarrow0$.

\subsubsection{Precision of estimation and its scaling with $R$ and $N$}\label{sec:dimconv}

Considering that e.g. $\hat{\lambda}^{(T)}_r$ is defined similarly to $\hat{\kappa}^{(T)}_r$, its variance scales similarly as given by Eq. (\ref{eq:ftescscaling}), which is inherited, along with the Gaussian form of limit distributions, by $\hat{D}_{KGa,r}^{(2,T)}$. Considering averaging with respect to realization too, we have: 
\begin{equation}\label{eq:DKGscalingn}
    \sigma^2_{\hat{D}_{KGa}} \propto \frac{1}{TR}.
\end{equation}
For some fixed $R$, using relationship (\ref{eq:memory}), we have then:
\begin{equation}\label{eq:nonnormal}
    \sigma^2_{\hat{D}} \propto \ln1/\varepsilon_* \propto 1/\ln N.
\end{equation}
We note first that these scaling laws of precision apply to attractors of closed systems too, in agreement with previous derivations by Namenson et al. \cite{NOA:1996}. Second, they are formally analogous to scaling laws of the escape rate estimates, Eqs. (\ref{eq:ftescscaling}) and (\ref{eq:ftescscalingN0}), respectively. Therefore, third, just like with the escape rate, it would be an overwhelming numerical burden to improve the precision of the dimension estimate by increasing the ensemble size. It is done much more effectively by producing a number of estimates with different realizations of the perturbation sequence and averaging them. However, unlike in case of the escape rate, this approach for estimating the dimension introduces a bias, as discussed previously in Sec. \ref{sec:averaging}.

\subsection{Examples}\label{sec:dimexamples}

We consider here the two examples specified in Sec. \ref{sec:escexamples}.

\subsubsection*{Example 1: Area-preserving naturally open baker map}

\begin{figure}
\centerline{\psfig{figure=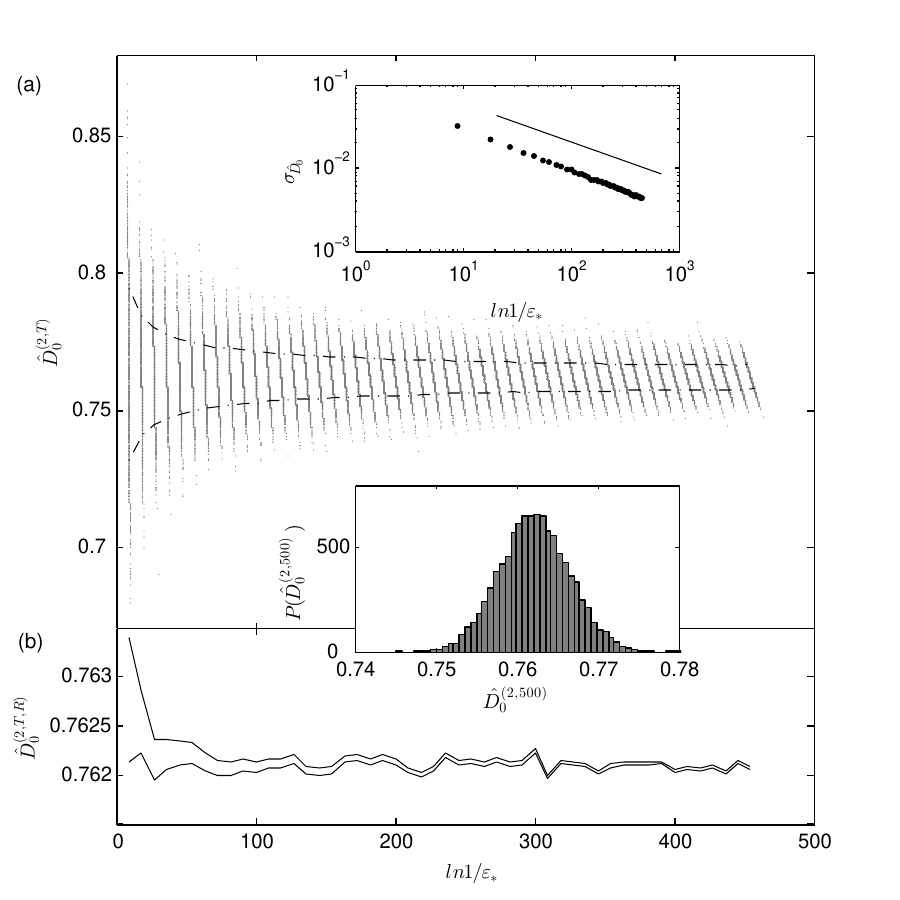,width=\linewidth}}
\caption{Fractal dimension in the area-preserving naturally open baker map. (a) Fractal dimension $\hat{D}_{0}^{(2,T)}$ vs minimal box size $\varepsilon_*$. A number of $M=10^4$ experiments (realizations of $a_t$) were carried out to generate statistics. The standard deviation of $\hat{D}_{0}^{(2,T)}$ with $T$ fixed is marked by a pair of dashed lines. Lower inset: histogram of $\hat{D}_0^{(2,T=500)}$; upper inset: scaling of the dispersion $\sigma_{\hat{D}_0}$ (half the vertical space between the dash-dot lines, approximately) as $(\ln1/\varepsilon_*)^{-1/2}$ for large $\ln1/\varepsilon_*$'s, the $\varepsilon_*$'s belonging to the respective expected values $\mu_{\hat{D}_0}$ (the solid line has a slope of 1/2). (b) Comparison of the harmonic (lower line) and arithmetic mean (upper line) estimates. For both panels $T=10l,\ l=1,\ldots,50$. Discrete data points which indicate the means and standard deviations of dimension estimates are connected to guide the eye.} \label{fig:D0vseps01} \end{figure}

By this example we intend to support the claim that dimension estimates are generally biased, and also the scaling laws of the precision of estimates given by Eqs. (\ref{eq:memory}) and (\ref{eq:nonnormal}). The finite-time positive Lyapunov exponent can be obtained such as: $\hat{\lambda}^{(T)}_r = \langle\ln a_{t,r}\rangle_{t=1}^T$. Note that after $T$ iterations each stripe is stretched by a factor of $\Pi_{t=1}^Ta_t$. 
On the other hand, having already obtained an analytic expression for $\hat{\kappa}^{(T)}_r$ given by Eq. (\ref{eq:bakerescrand}), we can write the order-$T$ approximant to the dimension across (along) the unstable (stable) manifold as: 
\begin{equation}\label{eq:OnapproxD0}
    \hat{D}_{KGa,r}^{(2,T)}=\hat{D}_{0,r}^{(2,T)}=
    \frac{\ln2}{\langle\ln a_{t,r}\rangle_{t=1}^T}.
\end{equation}
We numerically generate a large $M$ number of sequences of $a_t$ of length $T$, by which the first part of scaling law (\ref{eq:nonnormal}) can be prompted as follows. (For simplicity $R=1$, and we omit the realization index $r$ in the notation.) The width of the stripes can be considered as the minimal box size to cover the order-$T$ approximant of the manifold: $\varepsilon_*=\Pi_{t=1}^T 1/a_t$. Thus, not only the dimension estimate $\hat{D}_{0}^{(2,T)}$ depends on the realization of the random sequence $a_t$, but also the minimal box size $\varepsilon_*$. Therefore, it is not straightforward to generate a distribution of $\hat{D}_{0}^{(2,T)}$ belonging to the same $\varepsilon_*$ just by generating a large $M$ number of `realizations'. However, by increasing the sequence length $T$, the effects of decreasing box size can be explored. In Fig. \ref{fig:D0vseps01} (a) dimension estimates $\hat{D}_{0}^{(2,T)}$ are plotted against corresponding minimal box sizes, or rather $\ln1/\varepsilon_*$. For each fixed $T$ the data points align to a hyperbola segment. The large number of realizations generate sampling distributions of $\hat{D}_{0}^{(2,T)}$ and $\varepsilon_*$ or $\ln1/\varepsilon_*$. The vertical distance between a pair of dash-dot lines indicate the standard deviation of the sampling distribution of $\hat{D}_{0}^{(2,T)}$. The upper inset in the figure shows the standard deviation $\sigma_{\hat{D}_0}$ against $\ln1/\varepsilon_*$, belonging to fixed $T$'s, and confirms that scaling law (\ref{eq:nonnormal}) holds. The lower inset shows the sampling distribution of $\hat{D}_{0}^{(2,T=500)}$, which can be well-approximated by a Gaussian form.

We can obtain the scaling laws formally as well, as follows. We shall introduce now the following notation involving the minimum box size: $\nu=\ln1/\varepsilon_*$, and for the baker map we have that: $\nu=T\langle\ln a_t\rangle$. It follows then that: $\sigma_{\nu}^2\propto T^2/T=T$. Next, Eq. (\ref{eq:OnapproxD0}) can be rewritten as: $\hat{D}_0=\ln 2T/\nu$. If we linearize this about the mean $\bar{\nu}$, we can establish that: $\sigma_{\hat{D}_0}^2\approx |\hat{D}'_{0,\nu}(\bar{\nu})|^2\sigma_{\nu}^2$. That is, in the large-$T$ limit: $\sigma_{\hat{D}_0}^2\approx\ln^22T^2/\bar{\nu}^4T$. From the definition of $\nu$ we have that $\bar{\nu}\propto T$, and with this: $\sigma_{\hat{D}_0}\propto1/\bar{\nu}$, which conforms with Eqs. (\ref{eq:DKGscalingn}) and (\ref{eq:nonnormal}). 

We note that in this special case, considering Eq. (\ref{eq:OnapproxD0}), the harmonic mean of the $\hat{D}_{KGa,r}^{(2,T)}$ (or $\hat{D}_{BC,r}^{(2,\varepsilon_*)}$) values would provide an (approximately) unbiased estimator of $\hat{D}_1^{(2)}=\hat{D}_0^{(2)}$. A comparison of the harmonic and arithmetic means is shown in Fig. \ref{fig:D0vseps01} (b), where their difference for decreasing $T$ or $\ln1/\varepsilon$ is clearly indicated, which gives the measure of the bias of the arithmetic mean.

\subsubsection*{Example 2: Area-preserving closed baker map with a leak}

\begin{figure}
\centerline{\psfig{figure = 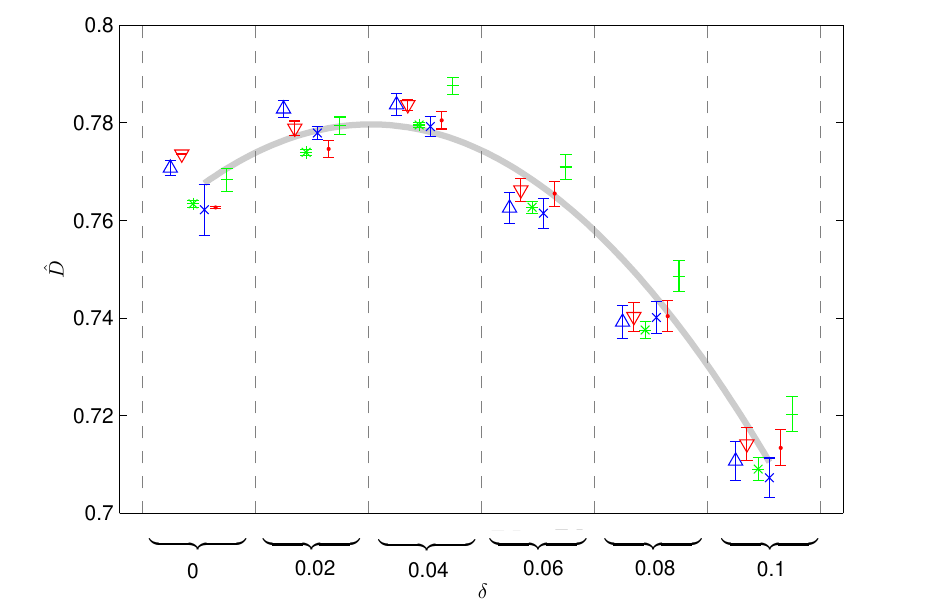,width=\linewidth}}
\caption{(Color online) Fractal dimensions in the area-preserving closed baker map with a leak for several values of the additive perturbation strength $\delta$. Errorbars indicate the standard deviation of the sample mean in the spirit of Eq. (\ref{eq:DKGscalingn}). An increasing number of experiments were done for increasing $\delta$ ($R=5\cdot10^3\times\delta$, and $R=10$ for $\delta=0$). Five different estimators are employed as described in the Appendix. All estimates are obtained for six values of $\delta\in[0,0.1]$ with 0.02 increments, but results using the different estimators for the same $\delta$ are plotted with a spacing for better visibility. A gray curve corresponding to the best fitting quadratic polynomial emphasizes the nonmonotonic dependence of the dimension on the perturbation strength.
} 
\label{fig:dimvarysigleakadd}
\end{figure}

The fractal dimension $\hat{D}_0^{(1)}=\hat{D}_0^{(2)}$ is evaluated for the same values of the perturbation strength $\delta$ as considered previously in case of evaluating the escape rate (Fig. \ref{fig:kappavsigvert} (b)). It has been estimated in five different ways using the estimators defined in Sec.~\ref{sec:averaging}, distinguished by different markers in Fig. \ref{fig:dimvarysigleakadd}. (The reader is referred to the Appendix for details.) The results indicate a characteristic enhancement of fractality, with a maximal value of $\hat{D}$ for some finite $\delta$. This value should be the same as the one for which maximal trapping occurs, for the following reason. With additive perturbation of the baker map, the Lyapunov exponents are unchanged, which is not simply because the perturbation terms do not explicitly appear in the Jacobian matrix, but also because the Jacobian matrix is the same in every point of phase space, for which reason the randomly perturbed trajectories do not have an influence either. Thus, considering the KG relations (\ref{eq:KGrand}) for random maps, the dependence of the dimension on the noise strength is inherited solely from that of the escape rate {in this example. If the Lyapunov exponents monotonically depend on the noise intensity, which is believed to be the case in general, then noise-enhanced fractality and uncertainty should always accompany noise-enhanced trapping.}

The different estimators seem to be biased to different degrees, as the errorbars do not overlap systematically. The estimator $\hat{D}_{KGb}^{(2,T,R)}$ (marker $\times$) {has been shown to be unbiased and thus it is expected to yield the most accurate figures, closely approaching the true value. Therefore,} its mismatch with the other markers indicates the bias of the corresponding estimators. Since they group fairly closely around the true value, they are also capable of robustly indicating the nonmonotonic behavior.

\section{Conclusions} \label{sec:conclusions}

We have investigated how the characteristic measures of transiently chaotic systems (e.g., $\kappa, D_0,\lambda$) depend on the type (noisy or random maps) and strength of the stochastic perturbations. Random maps are described in terms of a time-dependent conditionally-map-invariant measure, a measure which we introduced as a natural combination of concepts from transient chaos theory \cite{DY:2006,LT:2011} and (dissipative) random maps~\cite{Arnold:1998,BKT:2011}. For any fixed time, this measure exhibits a clear fractal character, similarly as the measure of the autonomous system. As in the case of attractors~\cite{BKT:2011}, we argue that the measure of the noisy map corresponds to the average of the time-dependent random map measure over different times (or realizations) and is smooth on fine enough scales. Based on this description we showed that the escape rate for the random map~$\hat{\kappa}$ is always larger than the one of the noisy map~$\tilde{\kappa}$ (for the same perturbation strength). 

All measurements in numerical and experimental situations are limited to finite numbers of trajectories~$N$ and realizations~$R$. We have shown that in the random map the precision of finite-time estimates of $\hat{\kappa}$ and $\hat{D}$ alike converge extremely slowly with $N$, typically as $\propto 1/\ln N$. We have shown that this limitation can be compensated by estimating through averaging over different realizations $R$, whereby the precision scales as $\propto 1/R$. In the case of the fractal dimension $\hat{D}$, however, our results indicate that the different finite-size estimators are typically biased, and even inconsistent as $R \rightarrow \infty$. {This means that the quality of estimating $\hat{D}$, depending on $N$ and $R$, has to be carefully analyzed, to guarantee that the inaccuracy (bias) and precision (spread) are small.}

Our results regarding the dimension gain a {practical} meaning through its relation with uncertainty: the greater the dimension, the greater (smaller) the uncertainty (exponent $\alpha$), as can be seen by the relation $\alpha=1-D$ in Eq.~(\ref{eq:ucexpanddim}). The finite-size/resolution estimates of $\hat{D}$ vary with the realization of the stochastic perturbation process, as a consequence of which the uncertainty (due to the uncertain choice of the initial condition) depends on the specifics of the current perturbation too. 

{Similarly to noise-enhanced trapping~\cite{AE:2010}, in random maps we have observed a nonmonotonic dependence of $\hat{\kappa}$ and $\hat{D}$ on the perturbation strength $\delta$, when the extrema occur at approximately the same finite value of $\delta$.} This entails that for the same perturbation strength the uncertainty is also maximal. The intuitive picture for this is that perturbation-enhanced trapping increases the chaotic life time of the trajectory, which makes it less predictable. The perturbation strength have to be increased beyond this point to steadily improve predictability. 

{Finally, we consider a concrete physical situation in which our results could be tested. Consider a two-dimensional~$\vec{x}\in \mathbb{R}^2$ fluid flow {exhibiting} a velocity field $\vec{v}(\vec{x},t)$ with a complicated dependence on $t$ (or being stochastically perturbed), leading to a transiently chaotic dynamics of fluid particles $\dot{\vec{x}}=\vec{v}$ (see, e.g., Ref.~\cite{NT:1998} and references therein). We are interested in measuring the spatial evolution and lifetime of tracers in an observational region. We consider two experimental {protocols}: (i) single tracers are measured successively, and the results over different tracers are combined; (ii) an ensemble of tracers is used {in} each experiment. Identifying case (i) with the noisy-map scenario and case (ii) with random-map scenario, we predict for the lifetime of tracers an exponential decay with an escape rate  $\kappa_{(ii)} \ge \kappa_{(i)}$, and for the spatial pattern at a fixed time ({measured from the} placement of tracers) a fractal dimension $D_{(ii)} \le  D_{(i)}=2$.}

\section*{Acknowledgements}

Useful discussions with J. C. Leit\~ao and T. T\'el are acknowledged. {The authors would like to thank R. Klages for calling their attention to references [23] and [28] after the submission of the manuscript.} T.B. is grateful for a postdoctoral fellowship given by the Max Planck Society. This work was also supported by the Hungarian Science Foundation under grant number OTKA NK100296.

\section*{Appendix}

\section*{Numerical computation of the $\hat{D}$ estimators}

Results distinguished by different markers in Fig. \ref{fig:dimvarysigleakadd} correspond to the different algorithms described in Sec. \ref{sec:averaging}, and further details are provided below. Before each description, the code name of the algorithm is followed by the type of marker used.

(I:BC.a: $\bigtriangleup$ and $\bigtriangledown$) In this special case of the area-preserving closed baker map with a vertical leak centered around $x=1/2$, the stable and unstable manifolds in 2D are aligned to straight horizontal and vertical lines, respectively. This allows for the collapse of all data points and box-counting in 1D. The patterns of both the stable and unstable manifolds are irregular when perturbations are present, meaning that different lines that constitute the manifold may have different lengths. With collapsing the unstable manifold, despite its irregularity as just described, the true value of the fractal dimension is not affected. Numerical results, however, reject equality of the dimension estimates for two values of the perturbation strength ($\delta=0$, 0.02). This may be because the irregular geometry of the manifolds introduces an additional bias of the very simple box-counting dimension estimator. For the simulation a number of $N=5\cdot10^6$ trajectories are initially uniformly distributed in $(x_0,y_0)\in[0,1]\times[0,1]$. The unstable manifolds are approximated by the trajectories which did not escape after $t_f=20$ iterations.

(I:BC.b: $*$) Fractality across the stable manifold can be resolved also by initializing the ensemble ($N=5\cdot10^5$) along a line across the manifold (e.g. $x_0=0.3$ and $y_0\in[0,1]$).

(II:KG.b: $\times$) For the estimator $\hat{D}_{KGb}^{(1,T,R)}$ the escape rate $\hat{\kappa}^{(T_{\kappa},R)}$ is evaluated based on the interval $t\in[6,t_f/2]$, where $t_f$ is the minimum of all simulation run times at which all trajectories are already escaped (same simulation as for III:$\alpha$). 

(III:$\alpha$.a: $\bullet$) The basin boundary is determined along the same line as for (I:BC.b), and the dimension of it is estimated by fitting the scaling line $\mathcal{N}_b(\varepsilon)$. For this the finest resolution is facilitated by $N=5\cdot10^5$ trajectories. In the case of the unperturbed baker map the same basin boundary is obtained by checking whether the left or right side of the leak-, or whether the lower ($y<b$) or upper ($y>b$) regime the trajectories escape from. For the results presented the first option was taken. This is the estimator that overall best conforms with the supposedly most accurate estimation (II:KG.b).

(III:$\alpha$.b: $+$) The dimension can be obtained also by first estimating the uncertainty exponent by fitting the scaling line $\mathcal{N}_b(\varepsilon)/\mathcal{N}_0(\varepsilon)$ and then applying relation (\ref{eq:ucexpanddim}). Interestingly, this approach modifies both the accuracy (leading to overestimation) and precision of estimation.

\end{document}

%% file: table-measures.tex
\begin{table}
\begin{center}
\begin{tabular}{|l|| c | c | c |}
\hline
{}
&{autonomous maps}
&{random maps}
&{noisy maps}
\\\hline
\hline
{attractor}
&{strange}
&{snapshot}
&{noisy}
\\\hline
{measure}
&{SRB}
&{$t$-dep't sample-}
&{fuzzy}
\\
{}
&{$\mu^{att}$}
&{$\hat{\mu}^{att}_{t,r}$}
&{$\tilde{\mu}^{att}=\langle \hat{\mu}^{att}_{t,r} \rangle_r=\langle \hat{\mu}^{att}_{t,r} \rangle_t$}
\\\hline
\hline
{saddle}
&{fractal/chaotic}
&{snapshot}
&{noisy}
\\\hline
{measure}
&{c-measure}
&{$t$-dep't c-}
&{fuzzy c-}
\\
{}
&{$\mu$}
&{$\hat{\mu}_{t,r}$}
&{$\tilde{\mu} = \langle \hat{\mu}_{t,r} \rangle_r = \langle \hat{\mu}_{t,r} \rangle_t$}
\\\hline
\end{tabular}
\caption{Summary of measures of chaotic systems discussed in the text. The measure in association with the saddle is supported by the unstable manifold of it \cite{LT:2011}.}\label{tab.measures}
\end{center}
\end{table}

%% file: NoisestrengthRII.bbl
\begin{thebibliography}{25}
\bibitem{noreturn} We assume that trajectories that leave $\varGamma$ never return to it.
\bibitem{nohat} To simplify notation we avoid using a hat for $N(t)$ as was used for $\hat{\mu}$.
\bibitem{morefundamental} This limitation is more fundamental than being simply due to a limited simulation time; the memory property (\ref{eq:memory}) and scaling law (\ref{eq:nonnormal}) hold also in case of closed dissipative systems, when the simulation time is not restricted by $N$.
\bibitem{twoexperiments} Taking \cite{musamplevsc} into account, we note that from a computational point of view, corresponding values of $\hat{\kappa}^{(T)}_r$ and $\hat{\lambda}^{(T)}_r$ with respect to the same time window can be obtained from two separate experiments only, using the same realization of the perturbation. In one experiment $\hat{\kappa}_{t,r}$ values are selected from the end of a numerically obtained sequence, where trajectories approximate the unstable manifold, and in another experiment $\hat{\lambda}_{t,r}$ values are selected from the middle, where trajectories approximate the saddle. At the expense of inaccuracy, the ensemble average $\langle \hat{\lambda}_{t,r,n} \rangle_n$ to approximate $\hat{\lambda}_{t,r}$ can be taken in the same window of time when the $\hat{\kappa}_{t,r}$ values are collected, in a single experiment. We took this approach in our numerics.
\end{thebibliography}
